# Web Search Result Clustering based on Heuristic Search and k-means

Mansaf Alam and Kishwar Sadaf

**Abstract**

Giving user a simple and well organized web search result has been a topic of active information Retrieval (IR) research. Irrespective of how small or ambiguous a query is, a user always wants the desired result on the first display of an IR system. Clustering of an IR system's result can render a way, which fulfills the user's actual information need. In this paper, an approach to cluster an IR system's result is presented. The approach is a combination of heuristics and k-means technique using cosine similarity. Our heuristic approach detects the initial value of k for creating initial centroids. This eliminates the problem of external specification of the value k, which may lead to unwanted result if wrongly specified. The centroids created in this way are more specific and meaningful in the context of web search result. Another advantage of the proposed method is the removal of the objective means function of k-means which makes clusters' sizes same. The end result of the proposed approach consists of different clusters of documents having different sizes.

## 1. Introduction

Today's conventional IR systems like web search engines, give millions of documents in an answer to a simple query. Of course it is overly tiresome to go through all the pages or documents. If the query is ambiguous and small, search engines supply even more number of results in order to provide documents for different meaning of the query. But a user does not want to and cannot traverse the whole result set. In [1], authors point that users visit only first few result. Moreover the first display of the result is dominated by the pages or documents that are frequently searched. For example, if "black berry" is queried, search engines give result that is dominated by pages related to "blackberry phone". Ambiguous queries create problems not only for search engines but for users too as users have to filter search for the desired page in the huge result. If "puma" is queried, it does not only indicate the brand Puma. By "puma", a user may imply "a large cat or panther" or a language or a kind of knife. A user may get lost in the huge result set returned by the search engine. A way to organize such an enormous set of documents is to group them into different clusters where each cluster may signify one possible meaning of the query. Document Clustering emerges as a powerful technique which separates unrelated documents and groups documents containing same topic. The unsupervised feature of the clustering technique makes it perfectly applicable for the clustering of web search result documents. Closely related to web search result clustering (SRC) is Image Search Result Clustering (ISRC) in which images returned by an image retrieval system are clustered based on annotations and visual models [2]. The Scatter/Gather system by Cutting et al [3] is held as the conceptual father of all clustering engines. There are many commercial clustering engines available that clusters the web search results like Vivisimo, carrot2, kartoo, and duckduckgo etc [4]. Vivisimo is an interesting clustering engine but the underlying method has not been published yet. Carrot2 employs Singular Value Decomposition (SVD) method for clustering. Recently, information retrieval methods based on ontologies have been surfacing, but ontologies for each and every entity are not available on the web. In [5], authors present a method for clustering search result using ontologies. The query words are mapped to ontologies for possible categorization.

Clustering can be broadly classified as hierarchical and partitional or flat. These types of clustering can be applied to the search result. Hierarchical clustering can be further categorized as agglomerative and divisive [6], [7]. In agglomerative approach, initially each data object is assigned to different singleton cluster and then merged on some similarity criteria until a stopping criterion is met. Whereas in divisive clustering, all the data objects are assigned to one single cluster then divided into different clusters until a certain a criteria is fulfilled. Flat or partitional approach of clustering tries to group data objects in a single go where an objective function is minimized [8]. Hierarchical clustering's time complexity of $O(n^2)$ makes it unfeasible for web where speed matters. Flat or partitional clustering can meet the speed requirement, but bad initial choices of seeds can reduce the performance.

In the field of search result clustering research, many web features like hyperlinks, user's context or web usage etc. have been the topics of interest. Search result clustering can be classified as graph-based, rank-based or content-based [9]. However, most clustering engines provide results based either only on hyperlinks or on the topical similarity i.e. lexical similarity. The most common method employed by clustering engines is the clustering of short text or paragraphs returned with each result. These small paragraphs called snippets, give hint to what the actual documents contain. Suffix Tree Clustering (STC) [10] is a text based web search result clustering system (Grouper) where snippets of the resultant documents are clustered. The matching is done on phrases rather than one single word. Snippets are short paragraph usually two to three lines containing sentences which have keywords that appear in the query. Many studies [11, 12 and 13] have been based on STC. Instead of employing short snippet for clustering, we apply whole document. A short snippet cannot provide the whole outlook of the document and moreover if the snippet does not contain the exact similar words, it would not get into the appropriate cluster. [14] utilizes the whole document content rather than snippets in their clustering system.

In [15], authors provide a link based clustering method for web search result clustering using hyperlink analysis in co-citation and coupling. [16] presents a link based clustering of web search result where documents and links between them, are represented as graph. This study focuses on both aspects i.e. text as well as links. Hyperlinks are one of the important traits of the web documents. The documents or pages containing similar topic may or may not be hyperlinked. Authors in [17] propose a method of search result clustering based on heuristic search on the graph induced by the hyperlinks among the documents of search result. Another work which utilizes hyperlinks in heuristic search for the purpose of cluster labeling is proposed by [18]. Our work is based on the hyperlinks existing among the documents of the search result. The work presented here is a combination of heuristics and enhanced k-means method. By enhanced k-means, it is implied that k is not to be specified instead it is determined by the heuristics being applied. The specification of the factor k is known to be a drawback of the powerful k-means. The final result depends on this k. Wrongly chosen k tend to give off beam result. Many researches have been dedicated to estimate the initial value for k [19] and [20]. In [21], authors present an algorithm based on harmonic search and k-means and hold that even with bad initial choice of k, their algorithm performs well. K-modes and K-medoids are some variants of k-means technique. The similarity between documents is computed using Euclidean distance method in k-means technique. Other feature of k-means is the creation of equal sized clusters which is not suitable for document clustering in context of web search.

Spherical k-means [22] is another variant of k-means where similarity is measured through cosine similarity function. In our proposed work, we first find initial k centroids using heuristics and then assign documents to these centroids based on their cosine similarities. Cosine similarity measures the orientations of the documents instead of distance between them. For example two documents having word “apple” 500 times and 50 times respectively falls far apart when we measure their difference using Euclidean distance method but cosine similarity may find them having similar orientation. Our heuristic follows that in web search result, similar documents tend to link each other. This “tend to” relationship is strictly for web search context. We found that some similar documents or pages in “search result” share links. For example the query “puma” produces many pages. The pages that truly represent “brand puma” i.e. the pages from the puma brand merchandiser are connected. We used these linked documents to form the initial centroids using the heuristic. These centroids are topic specific thus produce quality clusters. The remaining documents, which do not share links, are cosine tested and assigned to appropriate centroids to form clusters. In the process new centroids may emerge if documents have less similarity with the existing centroids.

This paper is organized as follows. Section 2 presents a brief description of related techniques in the field of web search result clustering. In sub-section 2.1, a short review of k-means method is provided. In section 3, we elaborate our proposed method followed by dataset and experimental result in section 4 and 5 respectively. Evaluation of the proposed method is given in section 6. Page and document are used interchangeably throughout the paper.

## 2. Related Work

Clustering of web search result has been an interesting topic for research. Carpineto et al [3] gave an extensive survey that covers the popular clustering engines available on the Internet. As far as we know, this work is only the work that uses hyperlinks heuristic to determine the factor k to decide on the initial centroids for clustering. Another work based on heuristic search on the web search result graph is by Bekkerman et al [23]. This work is the only work which employs heuristic search in context of web search as far as we know. They employed heuristic search on the hyperlink graph of the web search result to prune unwanted edges and produce clusters. Our work differs from them as we consider only those hyperlinks which are present between the documents of search result without following links beyond the search result. Our heuristic search method resembles the beam search method. One feature of the beam search method is the requirement of specifying the width of the beam by the user. The width in our heuristic search is determined by the promising pages which have high connectivity. In [24] authors proposed a method to maintain monotonicity of beam search method by iteratively increasing the beam width. The proposed work here, considers the whole document for clustering rather than small snippets. Mecca et al [14] proposed a method to cluster search result using Singular Value Decomposition (SVD) on the whole document instead of snippets. Recently, in [25], authors propose a document clustering technique based on sampling, Particle Swarm Optimization (PSO) and k-means. There are many web search result clustering methods, [26, 27, and 28], but clustering in conventional search engines is yet to be achieved.

### 2.1 k-means

k-means is one of the fastest and common flat clustering techniques. It clusters each data point into one of K groups. K is a pre-determined positive integer that can be obtained by arbitrary selection or by some other training processes that observe the data relationships iteratively. k-means attempts to group data into k clusters by minimizing the mean-squared error

$$E = \frac{1}{n} \sum_{i=1}^{k} \|d - c_i\|^2 \quad (1)$$

where $c_i$ is the closest centroid to the data point $d$ and n is the total number of data points. If d is the nearest point to $c_i$, it is assigned to $c_i$ and then the mean is again computed. This function makes the centroids as compact and separated as possible. When the clusters are of varying sizes, k-means in order to reduce mean-squared error, divides large clusters into smaller ones even if they contain proper assigned data points. k-means method has the tendency to converge to local optima. To find a global solution, many methods are proposed. In [29], authors propose a fast global k-means technique based on geometrical information of the data points. Author in [30], relates the problem of assignment of points to centroid as black hole and star.

The classic k-means method uses Euclidean distance to measure the similarity between centroid and data points. Applying distance measure other than Euclidean may stop k-means from converging as its objective is to minimize the "squared" distance from centroid to data object. Spherical k-means, a variant of k-means employs cosine similarity method to compute the similarity between centroids and data points. All the centroids and data points are represented as vectors and L2 normalized. The cluster centroid is (L2) normalized summation of all the vectors in the cluster. Thus the centroid is also on the unit sphere. Zohng [31] observes that Euclidian and cosine methods give same result when applied on unit vectors.

## 3. Proposed Work

We propose an enhanced k means clustering approach where the factor k is not required to be specified externally and which produces varying size clusters. One of the drawbacks of k means technique is to specify k, prior to clustering. The final result of the clustering depends on this k. k-means uses the square-error method. In order to minimize the square-error, it splits large clusters even if they are well formed. It is not desirable. We hold that clustering is a domain dependent technique. The final clusters produced in a web search result clustering scenario, depend on the nature of the query and documents. So k means technique in its basic form would not be appropriate in this context.

We determine the value of k using a heuristic search on the result set. For better understanding, let's consider the pages or documents as nodes and links between them as edges. The heuristic we apply says: find a node with high number of links. If a node has many links to other nodes, implies that there are many edges between nodes. Clusters in a graph can be identified by high number of edges within and less number of edges between them. After applying heuristic we get k centroids based on the hyperlinks. In our proposed algorithm, to improve the understanding, we use the concept of agents that are assigned to each page of the search result. The function of each page's agent is to maintain the list of connected pages. Figure 1 algorithmically describes our proposed method.

Let the documents or pages be represented by $d_1, d_2, \ldots, d_n$ where $d_i \in D$ and $D$ is the result set consisting $n$ documents. Let $C_k$ represent the k centroids. At the end of the heuristic phase, we get $k$ different centroids $C_1, C_2, \ldots, C_k$ ie,

$$\bigcup_{i=1}^{k} C_j = \{d_1, d_2, \dots d_n\}\ and \quad C_i \cap C_l = \emptyset, i \neq l \tag{2}$$

To get initial *k* centroids, we first find the relation between the documents of the search result. To keep track of all the connected pages, each page $d_i$ maintains a list of pages that it reaches.

$$l(d_i) = \cup\, d_j \ \ , for\ all\ 1 \le j \le n\ and\ j \neq i \tag{3}$$

Input: A set of *n* web pages or documents, P= $\{p_1, p_2, \dots\dots\dots, p_n\}$

Output: A set of k partitions of documents, Q= $\{ q_1, q_2, .. q_k\}$, $q_i = \{ q_1, q_2, \dots q_n\}$ and $q_i \cap q_j = \emptyset$ , $i \neq j$

Step1: **Search result and heuristic**

Assignment of agents to each page

For each $p_i \in P$

Assign agent $a_i$ to $p_i$: $a_i(p_i)$

Initialize agent $a_i$'s promising page

$pr_i(a_i) \leftarrow p_i$

Initialize agent ai's list of connected pages: $L(a_i)$

**Searching for promising pages using heuristic**

For each page $p_i \in P$

Expand pis list of connected pages L

$L(a_i) \leftarrow L(pr_i)$

Filter L($a_i$) for new promising page

$pr(a_i) = p_j : argmax\|L(a_j)\|\ for\ all\ \ 1 \le j \le n , j \neq i\ and\ p_j \in L(a_i)$

Step2**: Centroid construction phase**

Initialize Singleton centroid $C_i \leftarrow Vector(p_i)$

Construct all pairs of pages $(p_i, p_{i'})\ s.t.$

$L(a_i) \cap L(a_{i'}) = \emptyset$

For each pair $(p_i, p_{i'})$

If $((p_i \in C_m) \cap (p_{i'} \in C_{m'})\ and\ C_m \neq C_{m'}$

Merge $C_m\ and C_{m'}$

$k = \|C_m\|\ for\ all\ m \in N$

Step3: **Assignment of document vectors to *k* centroids**

For each $p_i \in P$

Vectorize $p_i$:

$d_i \leftarrow p_i$

For each document vector $d_i$ , $1 \le i \le n$

Compute $d_i C_m \quad for\ m = 1,2, \dots, k$

Assign $d_i\ to\ C_m$ if

$C_m \leftarrow d_i\{y = argmax\ (d_i\ C_m)\}$ and

$y \ge \alpha$

Else create a centroid

$c_{k+1} = d_i\{y = argmin\ (d_i\ C_m)\}$

Re-compute the centroids

$C_m = \frac{\sum d_i}{\|\sum d_i\|}\ for\ all\ d_i \in C_m$

Figure 1: Algorithm for web search result clustering based on heuristic search and k-means

In the list maintained by each page, we try to find a promising page that links, to many pages of the search result set. We use the above mentioned heuristic. Where there is many links, clusters tend to form there. To find the promising page $p_i$ in a list, we search for the page which has the largest list.

$$p_i(d_i) = d_j : \arg max \ \left\| l(d_j) \right\| \ for\ all\ 1 \leq j \leq n\ and\ j \neq i \quad (4)$$

After obtaining the promising page $p_i$=$p_j$ for $d_i$, the list of $d_i$ now contains all the pages that list of $d_j$ contains. This process is performed until there is no change in the list of pages. The pages are merged into a centroid if their respective list contains same elements. At this point, we get *k* different centroids, which remove the requirement of specifying the factor *k* for k means method. To proceed further, all the documents and centroids are converted to vectors.

Let the document vectors be represented by $d_1,d_2,\ldots,d_n$, where each $d_i \in D$,and *D* is the search result set. The end result of the whole clustering process is k different clusters. Let $x_1,x_2,\ldots.x_k$ represent the *k* clusters such that:

$$\cup_{j=1}^{k} x_j = \{d_1, d_2, \ldots d_n\}\ and \quad x_j \cap x_l = \emptyset, j \neq l \quad (5)$$

The centroid of a cluster $x_j$ can be computed as

$$c_j \ = \sum_{d_i \in\, x_j} d_i / || \sum_{d_i \in x_j} d_i \ || \quad (6)$$

Where $c_j$ represents the centroid of $x_j^{th}$ cluster. To map document vectors to the relevant centroid, cosine similarity is used. The similarity between vectors can be easily interpreted by cosine method. The cosine similarity between document vector and a centroid vector can be defined as:

$$Sim(d_i, c_j) = \frac{d_i . c_j}{||d_i|| \, ||c_j||} \quad (7)$$

The similarity ranges from 0 to 1. The value 0 represents no similarity whereas similarity value 1 depicts complete similarity. This metric is a measurement of orientation and not magnitude. A document vector $d_i$ is compared to all the centroid vectors. $d_i$ is assigned to that centroid vector with which $d_i$ has the greatest similarity.

$$x_j = \{d_i : y = \arg\max d_i\ c_l\ \}\ for\ all\ l = 1,2,..,k\ and\ 1 \leq j \leq k \quad (8)$$

When a document vector gets assigned to a cluster centroid, the centroid needs to be updated.

$$c_j = \frac{\sum_{d_i \in x_j} d_i}{\left\| \sum_{d_i \in x_j} d_i \right\|} \quad (9)$$

Where $1 \leq j \leq k$.

The documents which have little similarities with centroids or similarity less than a given threshold, pose the problem of outliers. We get hold of this threshold value (α) after exhaustive computing of similarity between documents and centroids. Since we are clustering search result, a document cannot be left or thrown because of its less similarity with the existing clusters. It may be of importance to the user. To deal with the document vector which doesn't have similarity value greater or equal to the threshold value, we create a new partition and hold this document as a new centroid. If there is no further assignment of vectors to it, we hold it as a singleton cluster containing single document.

$$c_{k+1} = \{d_i : y = \arg\min d_i\, c_l\}, for\ all\ l = 1,2,..,k \quad (10)$$

Search engine gives multiple categories of documents in response to an ambiguous query. Many of these categories pertain to only one document of the search result. We relate this problem with our singleton clusters. Creating a group of dissimilar singleton clusters is less harmful than assigning them to well grouped clusters. One may think of this group of singleton clusters as "miscellaneous". The end result of the whole clustering process comprises of k clusters where each cluster contains similar documents which are dissimilar from documents of other clusters and a "miscellaneous" cluster.

## 4. Dataset and Experimentation

It is not feasible to assess the relevance of the huge result given by a search engine instead relevance is assessed for only a subset of the documents. The most standard approach is pooling [32]. In this system only top k documents returned by the search engine is assessed for relevance. In [33], authors proposed a pooling system where assessment of documents is done by assigning a score of importance to them. The documents are ordered by decreasing value of the importance score so as to provide less manual effort. We evaluated our method on two dataset containing web pages from first 10 and 15 display pages returned by Google in answers to the queries "jaguar" and "puma" respectively.

| Category | No. of pages | Category | No. of Pages |
|---|---|---|---|
| Car | 55 | Hotel | 2 |
| Animal | 17 | Photo Gallery | 1 |
| Sport | 4 | Movie | 1 |
| Super Computer | 1 | Touring | 1 |
| Scientific Prog. Package | 1 | Resin Models | 1 |
| Music | 1 | Timing Systems | 1 |
| Music Band | 2 | Eyewear | 1 |
| Computer Game | 1 | Financial Firm | 1 |
| Telecommunication Corp | 1 | Under Water Vehicle | 1 |
| Emulator | 1 | Mining | 1 |
| Magazine | 1 | | |

Table1: Jaguar dataset

| Category | No. of Pages | Category | No. of Pages |
|---|---|---|---|
| Puma Brand | 45 | Urban transport type | 1 |
| Web Server | 4 | Hotel | 4 |
| Animal | 17 | Unmanned Aircraft | 1 |
| Chocolate Puma Music Band | 4 | Energy fuel company | 1 |
| Darts | 1 | Golf | 1 |
| Brand Building | 1 | Dictionary | 1 |
| Helicopters | 3 | Marshal Art | 1 |
| Music Band | 1 | Medical Association | 1 |
| Bio Info Software | 1 | Travel trailers | 1 |
| Micro-array database | 1 | Program and Model analysis | 1 |
| Award | 1 | Blog | 1 |
| Puma Music | 1 | Club | 1 |
| Photo gallery | 2 | Game | 1 |
| Meteorological Data project | 1 | Pediatric medical | 1 |
| Tractors | 1 | Real estate | 1 |
| Intel Cable Modems | 1 | Tattoo | 1 |
| Financial site of puma brand | 1 | Knives Company | 1 |
| Vineyards | 1 | Speakers | 1 |
| Comic Character | 1 | Sports | 2 |
| Cycles | 1 | Biotech Firm | 1 |
| Weapon | 1 | Genes Binding | 1 |
| Steel Company | 1 | Multipole Algorithm | 1 |
| Movie | 1 | Virtual Pet Game | 1 |
| Movers and packers | 1 | Wines | 1 |
| Software | 1 | Car | 3 |
| Racing | 1 | Author | 1 |
| Mining | 1 | Chef | 1 |
| Air Compressor & tools | 1 | Lodging | 1 |
| Puma Biotech Inc. | 1 | Social Media training | 1 |
| Person | 1 | Bike | 1 |
| Bluetooth | 1 | Swimming Club | 1 |
| Fishing Lodge | 1 | Aircraft | 1 |
| Community Development | 1 | Distributor Company | 1 |
| Tools and Machines | 1 | | |

Table 2: Puma Dataset

There are 21 and 68 different categories in jaguar and puma dataset respectively.

## Preprocessing

Before preprocessing, we crawled these documents for direct hyperlinks. All the documents are first scraped for only normal text by excluding all the html and related tags. Usually preprocessing of a document involves tokenization, stop word removal and stemming. In tokenization, each word is tokenized. A token is a basic lexical unit of a language. Stop words are the words (for example “a”, “an”, “or” etc.) that occur frequently throughout the document set but do not convey any meaning. These words are removed in preprocessing.

In stemming, words are reduced to their root form. For example words “computer”, “computation” and “computing” share the same root word “comput”. We used Porter’s Stemmer algorithm for this purpose. Figure 2 shows a general preprocessing approach.

All these steps have been performed using Python language libraries. While extracting text from web pages, contents under the <meta> description tags, which are not part of the content but describe the contents are also extracted as there are many web pages that contain pictorial description of the topic with minimal textual information.

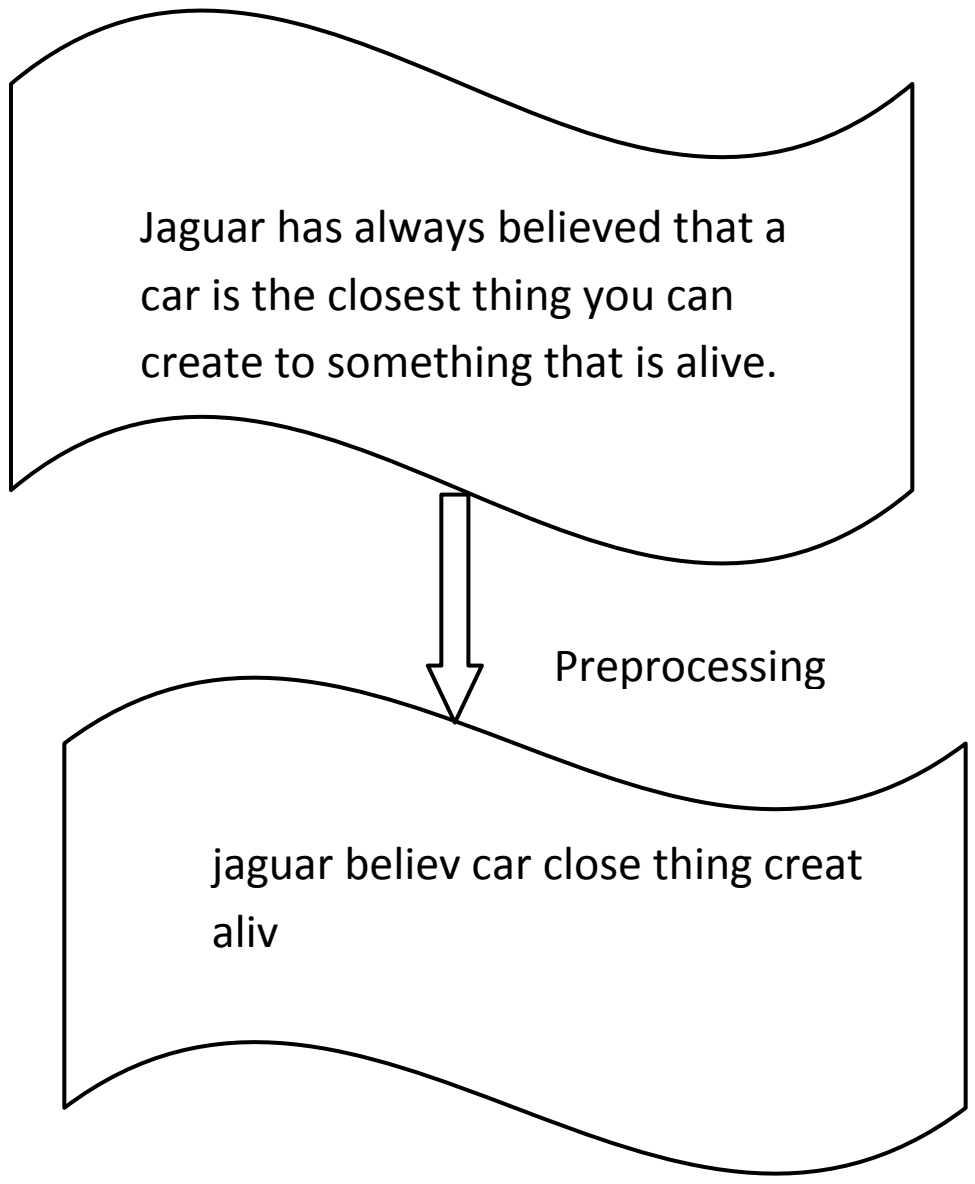

Figure 2: Preprocessing of Documents

**Document Vector and Cosine Similarity**

After obtaining k centroids, we changed the whole dataset and the centroids (tf-idf vectors) into unit vectors ($\|d\| = 1\ and\ \|C\| = 1$). We employed cosine similarity as it measures the orientation of the documents not magnitude. This orientation is angle between two vectors. Two vectors pointing far from each other could still have a small angle. In [34], authors state that cosine similarity gives better result.

For each document vector, cosine similarity is measured with every centroid. If the similarity to a centroid is greater or equal to a given threshold (α) of 0.50, the document is assigned to that centroid. If a document is not enough similar to the existing centroids, another centroid is created (k+1) with this document as the centroid concept. After exhaustive experimentation with the similarity measure, we found that documents with similarity below 0.50 are not related to or contain very minimal related contents, for the concerned centroid.

## 5. Result

There are techniques for cluster validation but a general framework for overall cluster validation is needed [33]. For the sake of simplicity, we concentrated on the large clusters that are identified by our clustering method. For jaguar dataset, we have car, animal and sport categories. Puma dataset provides four major categories of brand, mountain lion, web server and puma chocolate music band.

After applying heuristic search of our method on the dataset, we obtained 3 initial centroids in jaguar dataset containing nearly one third pages of the whole result.

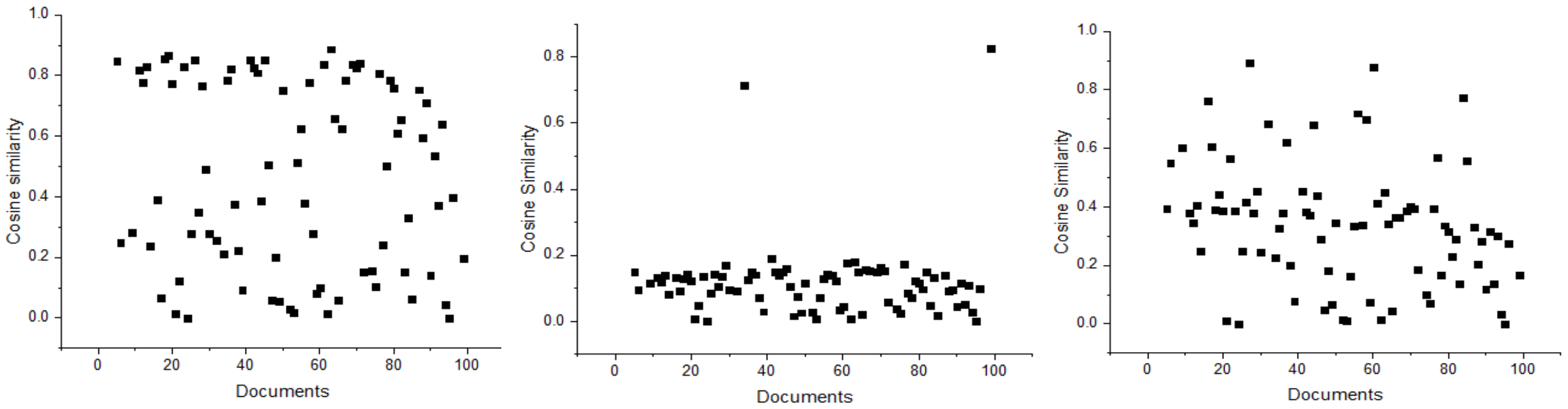

(a) Jaguar Car (b) Jaguar Animal (c) Jaguar Sport

Figure 3: Assignment of Jaguar documents based on cosine similarity

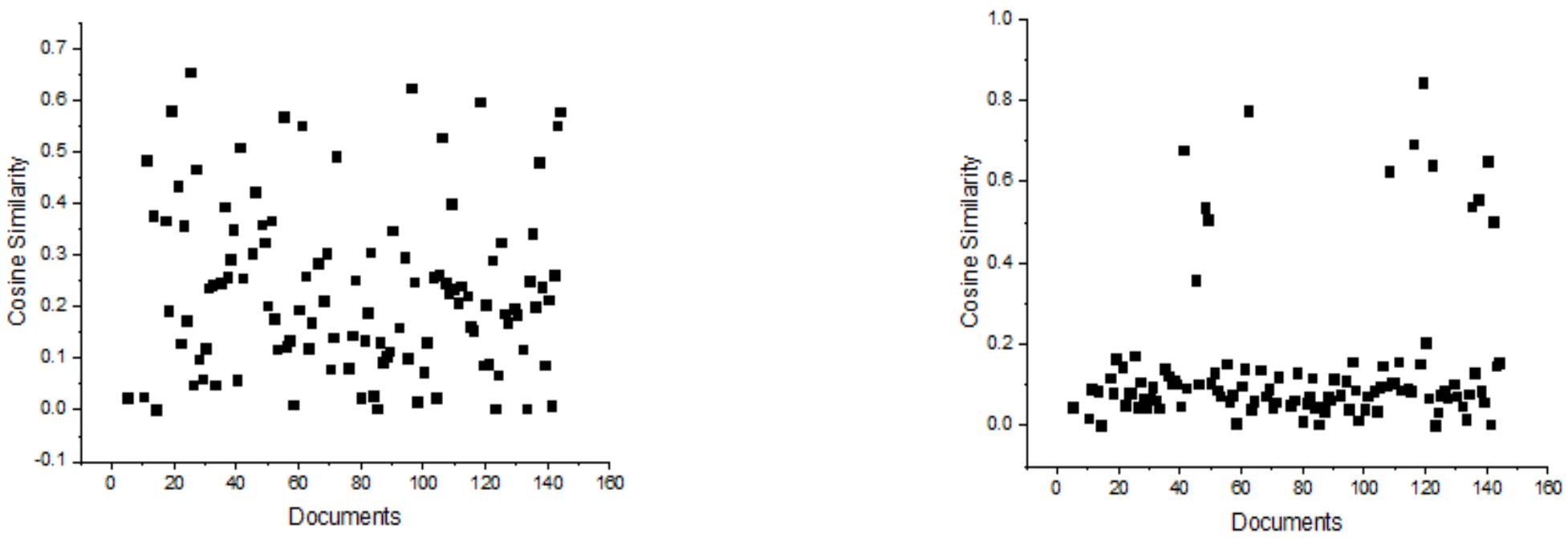

(a) Puma Brand (b) Puma-Mountain Lion

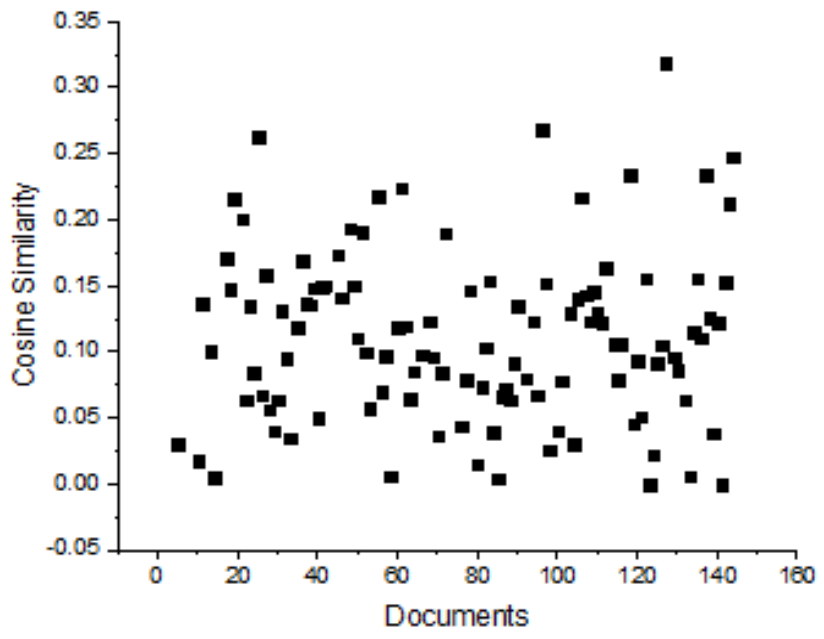

(c ) Puma Music Band

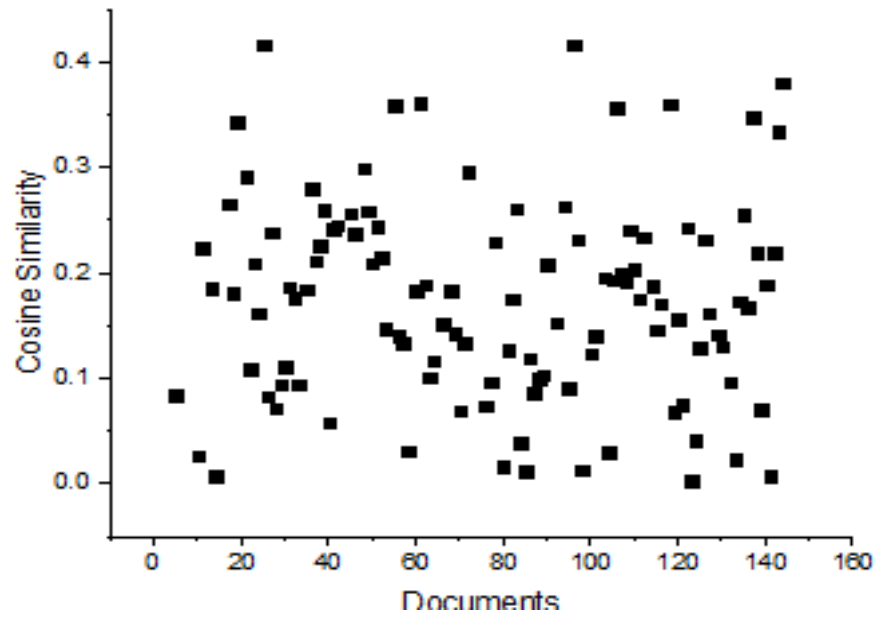

(d) Puma Web Server

Figure 4: Assignment of Puma Documents based on cosine similarity

Each figure represents the whole cluster. In Figure 3(a), documents are cosine tested with the car centroid. All the documents that have similarities greater that 0.50 are assigned to car centroid. Figure 3(b) represents the documents that are more similar to the sport centroid. The two documents in (b) are about Jacksonville Jaguar sports and contains only sport related contents. That is why we see a distinctive similarity. The documents in figure 3(c) depict the similarity with cat category cluster. Figure 4 represents the cosine similarity graph of 4 categories of puma. Around one third of clustered documents get assigned to the centroids in the heuristic phase of centroid creation.

To further support our method, we applied it on 1028 web pages related to smart phones. To eliminate the process of manually categorizing these many pages to evaluate our clustering method, we fired several very clear and specific queries. When specific and non-ambiguous queries are given to the search engine, the engine responds with result wherein first few pages are highly relevant to the query. In this fashion, we fired 27 queries to have 27 different categories. We feed all these web pages to our system without letting it know that these pages are results of different queries.

| **Category** | **No. of Pages** | **Category** | **No. of Pages** |
|---|---|---|---|
| Acer Liquid E2 | 47 | Blackberry Z30 | 46 |
| Celkon A40 | 46 | Dell Streak 7 | 49 |
| Gionee Elife E7Mini | 34 | HP Pre 3 | 49 |
| Huawei u8860 | 45 | Iball Andi 4 | 46 |
| Intex Aqua i7 | 48 | iPhone 4S | 19 |
| Jolla Phone | 42 | Lava Iris Pro 30 | 46 |
| LG Env Touch | 44 | LG Optimus G | 19 |
| Lumia 1520 | 45 | Micromax Canvas 4 | 16 |
| Motorola Moto G | 19 | Nexus 5 | 45 |
| Nokia X | 47 | Onida i666 | 43 |
| Qmobile | 46 | Samsung Galaxy Grand 2 | 17 |
| Samsung Galaxy S4 | 17 | Sony Xperia C | 18 |
| Spice SmartFlo Pace 3 | 45 | Videocon A53 | 46 |
| Xolo A500s | 46 | | |

Table 3: Smartphone Database

After applying our method, we found all the specific clusters related to above mentioned categories (Table 3). The hyperlink pattern among them is same as in our other two datasets. To see what each cluster holds, we extracted most frequent words in it. Figure 4 presents the most frequent words of the clusters. We used the “Wordle” online to get the layout. We provide Wordle with few most frequent words occurring in each final centroid. A picture speaks a thousand words. The boxes of Figure 5 represent most frequent words of various clusters with large and bold text to give a glimpse of the contents of the clusters.

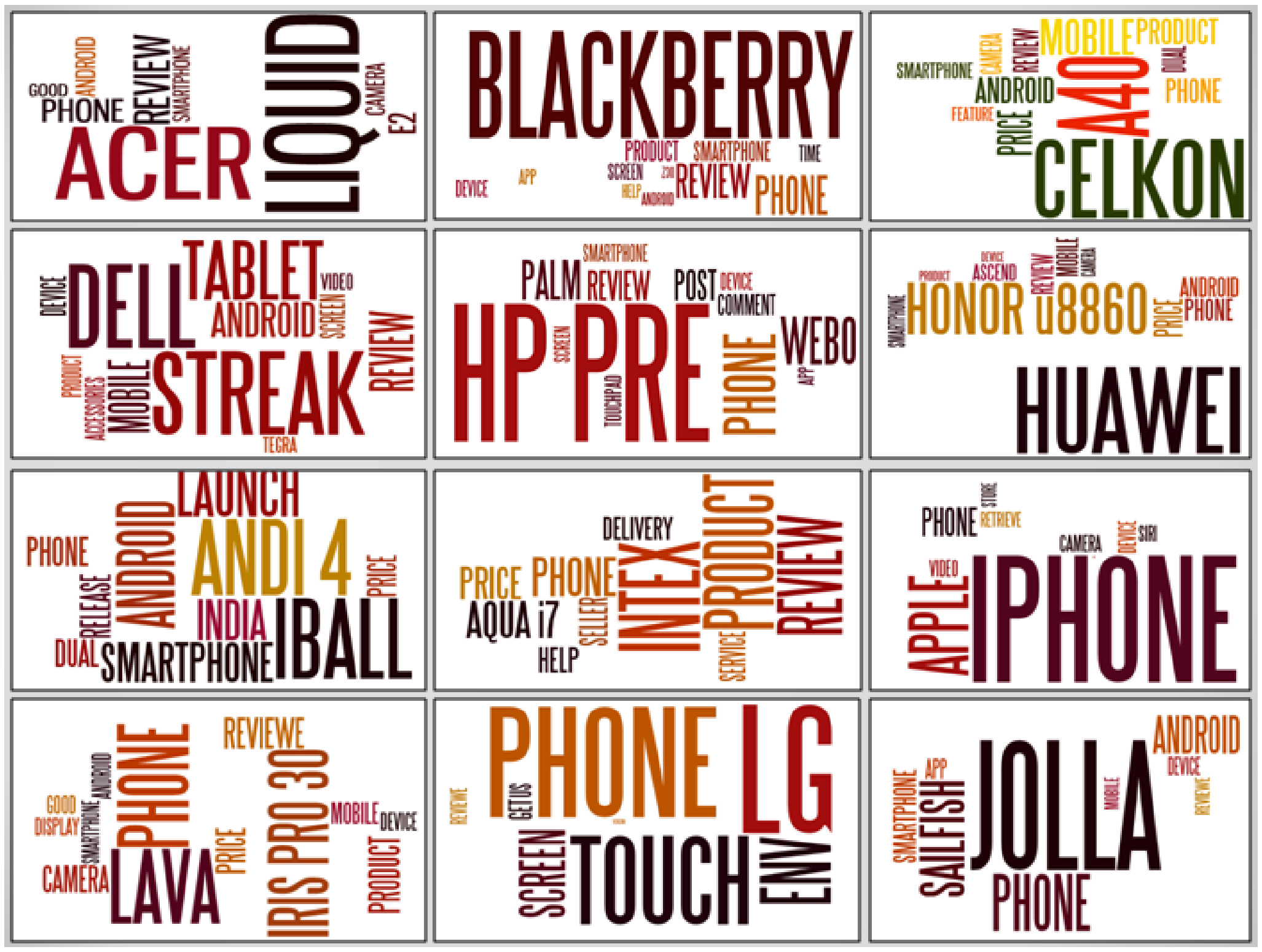

Figure 5: Prominent words of the clusters

## 6. Evaluation

To evaluate our method, we manually checked all the retrieved web search result pages obtained through pooling for topic classification. Although this process is time taking but to rightly evaluate the result, pre classification of the web documents is necessary. In Jaguar dataset, we classified 21 categories out of 100 web pages. In Puma dataset, 68 categories have been classified. Car, animal and sport topics emerged as major class in jaguar dataset. In puma dataset, brand, mountain lion, web server and music band emerged as main classes.

We evaluate our clustering method using Purity, Entropy, Precision and recall. These are standard measures for cluster qualities in the field of clustering and IR. Purity is the measure of coherence of a cluster. It measures the quality of a cluster by assessing how many documents are from a single category. Purity can be defined as:

$$P(C_i) = \frac{1}{n_i} \max_{\mathrm{h}}\left(n_i^h\right) \tag{11}$$

Where $n_i$ is the length of cluster $C_i$ , $\max_{\mathrm{h}}\left(n_i^h\right)$ is the number of documents that are from the leading category in $C_i$. Purity value of 1 means that cluster only contains document from single category.

Entropy is another method to assess the quality of a clustering method. It measures the distribution of classes in a cluster. Entropy value of 0 implies that the cluster consists of documents of only one class or category. While value 1 depicts that cluster contains a mixture of documents from different categories. It can be defined as follows:

$$E(C_i) = -\frac{1}{\log c} \sum_{h=1}^{k} \frac{n_i^h}{n_i} \log\left(\frac{n_i^h}{n_i}\right) \tag{12}$$

Where *c* is the number of categories in the dataset.

Precision and recall are another two methods to assess the quality of retrieved result in the field of IR. Precision is the ratio of the number of relevant documents retrieved to the number of documents retrieved and recall is the ratio of total number of relevant documents retrieved to the total number of relevant documents in the collection. A good retrieval system should have high precision for most levels of recall [35]. In the context of web search result clustering, each and every cluster is treated as different result set returned by the system in response for a query and each category is considered as the corresponding class. Precision P and Recall R are calculated as:

$$P = \frac{D_r}{D_t} \tag{13}$$

$$R = \frac{D_r}{D_{nr}} \tag{14}$$

Where $D_r$ ,is the number of relevant documents retrieved in a cluster, $D_t$ is the number of documents retrieved in a cluster and $D_{nr}$ is total number of relevant documents or number of documents in the corresponding class or category.

| Cluster # | Purity | Entropy | Precision | Recall |
|---|---|---|---|---|
| Cluster1 | 0.962 | 0.0811 | 96% | 93% |
| Cluster 2 | 0.944 | 0.0932 | 94% | 94% |
| Cluster 3 | 1 | 0 | 100% | 100% |

Table 4: Purity and entropy on 'Jaguar' dataset

| Cluster# | Purity | Entropy | Precision | Recall |
|---|---|---|---|---|
| Cluster 1 | 0.953 | 0.096 | 95% | 91% |
| Cluster 2 | 0.937 | 0.101 | 93% | 88% |
| Cluster 3 | 1 | 0 | 100% | 100% |
| Cluster 4 | 1 | 0 | 100% | 100% |

Table 5: Purity and entropy on 'Puma' dataset

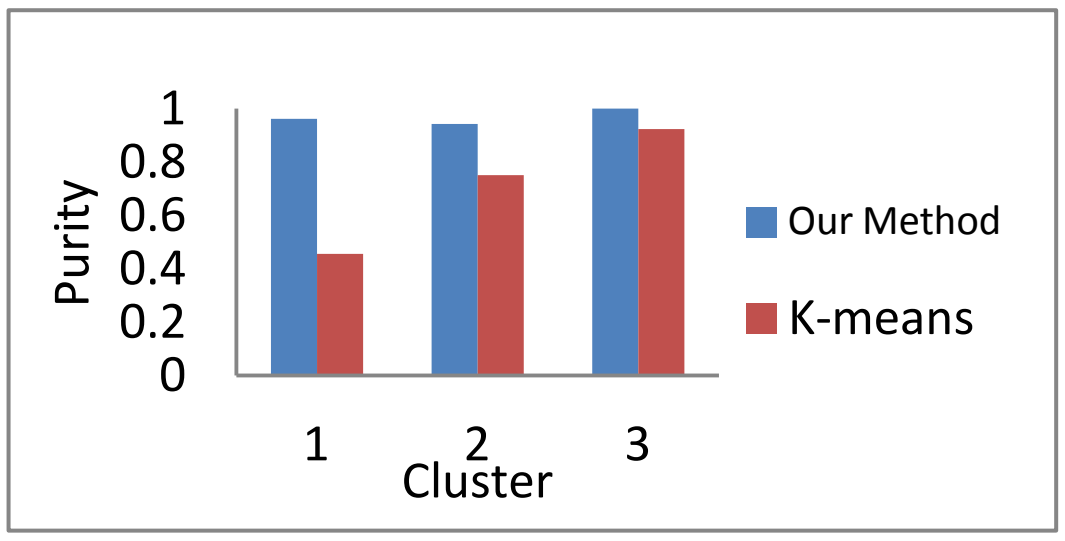

(a)Jaguar

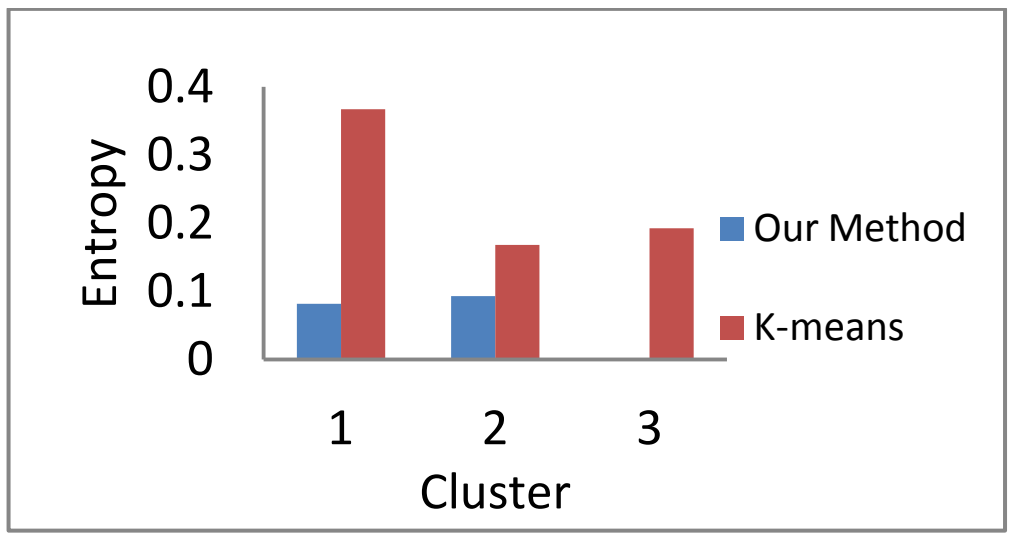

(b) Jaguar

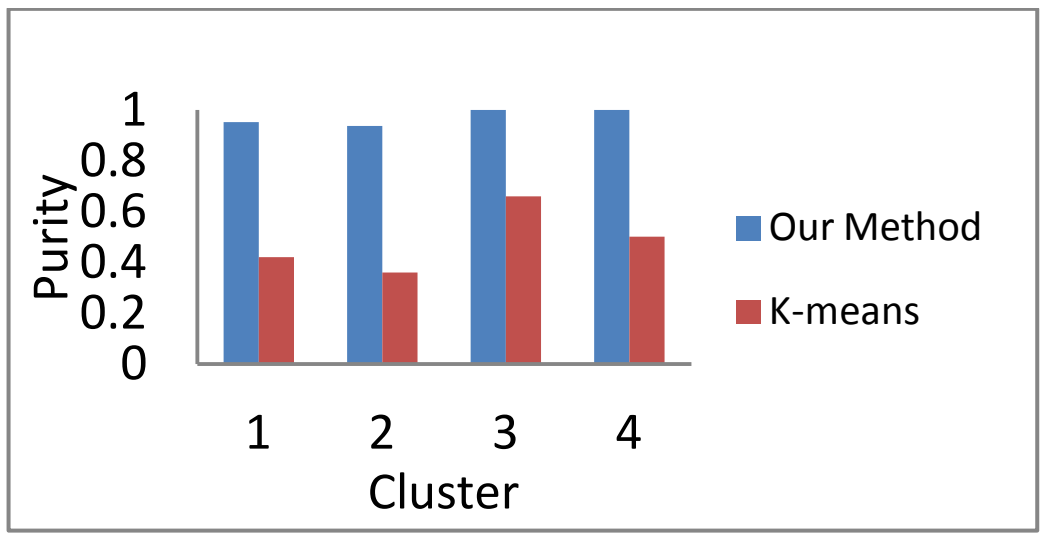

(c) Puma

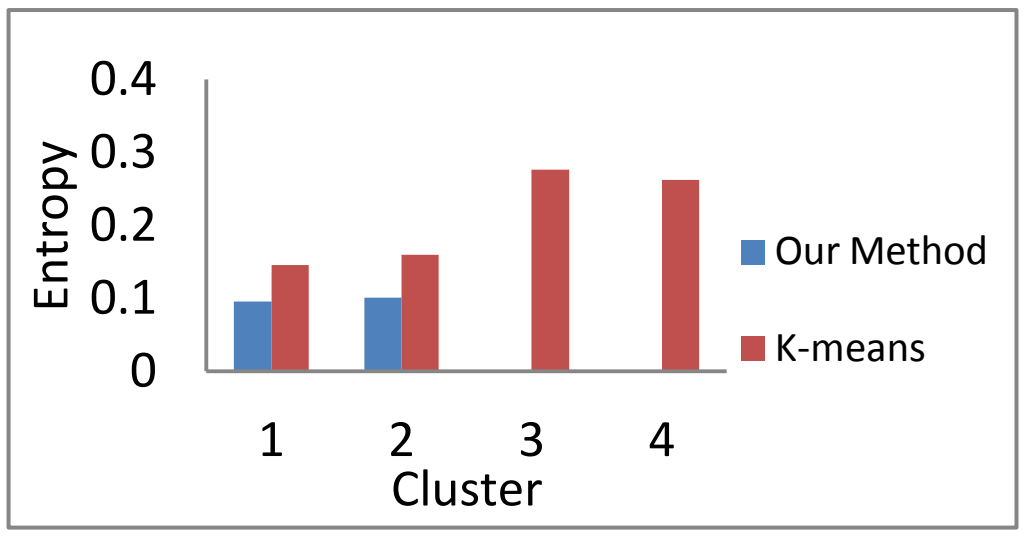

(d) Puma

Figure 6: Comparison of our method with k-means technique based on Purity and Entropy

For the sake of simplicity, we considered the largest categories of the jaguar and Puma datasets to evaluate our method. These categories are car, animal and sport in jaguar dataset and brand, mountain lion, web server and music band in puma dataset. Table 4 and 5 present the above mention evaluation metrics values for jaguar and pumas' largest clusters. We also compared our method with simple k-means method. Figure 6 presents the comparison of our method with the simple k-means method based on purity and evaluation metrics. The performance of our method is better than k-means. Our method reaches maximum purity and entropy on some clusters.

To further evaluate our clustering method, we applied four constraints laid down by [36] to assess the quality of a clustering evaluation metric. These constraints are cluster homogeneity, cluster completeness, cluster size vs. quantity and rag bag. Instead of evaluating above mentioned evaluation metrics like purity and entropy etc., we directly apply these constraints on our clustering method.  Cluster homogeneity states that a cluster should contain only similar documents. Cluster completeness specifies that documents from same class should be grouped in a cluster. Another constraint is cluster size vs. quantity which states that it is acceptable to have a small error in large cluster than having large number of small errors in a small cluster. These constraints are satisfied by our method by gaining high purity and low entropy values. Rag bag constraint holds that bringing dissimilar document to a clean cluster is damaging than having a cluster of dissimilar documents. Our method overcomes this constraint by creating a cluster of documents which are not similar to any centroids.

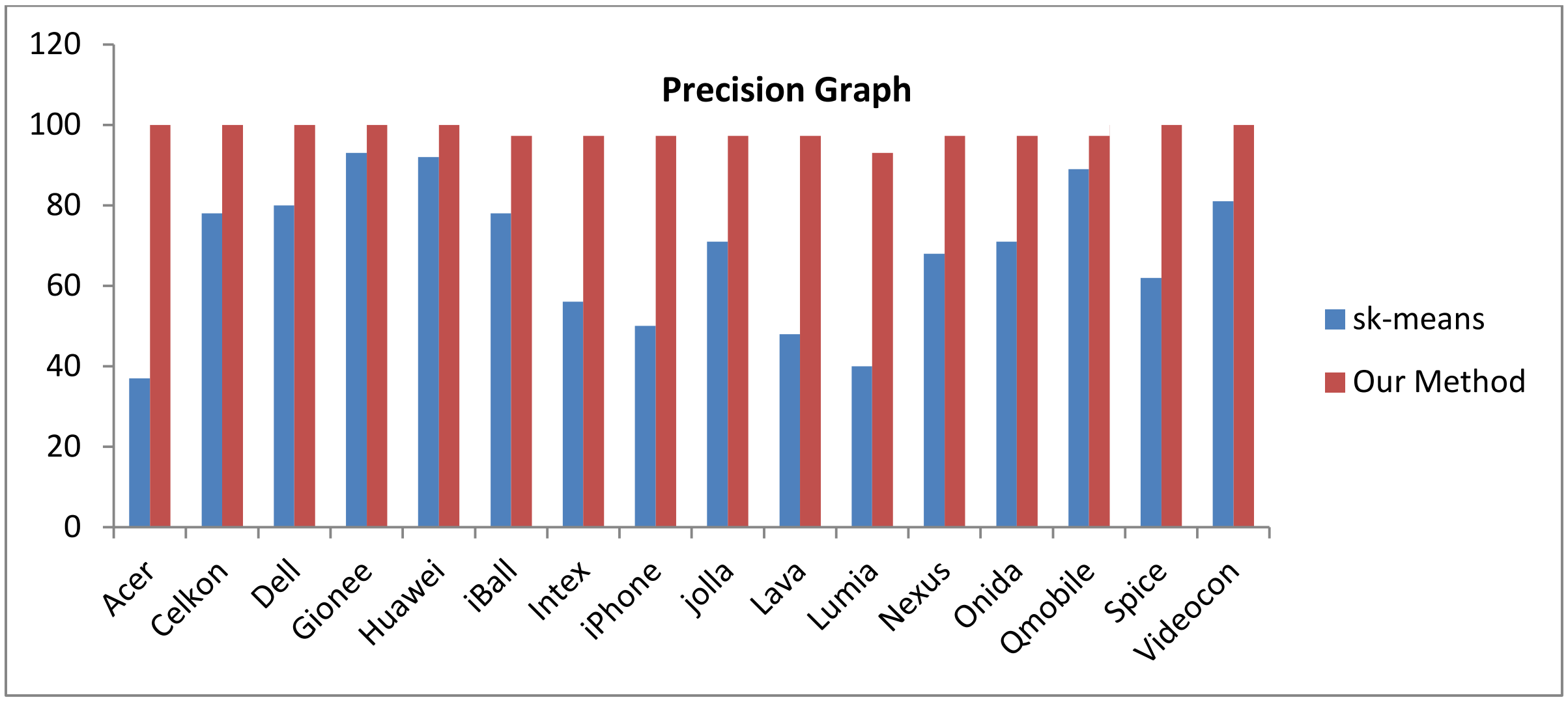

Figure 7: Precision values gained by our method and sk-means

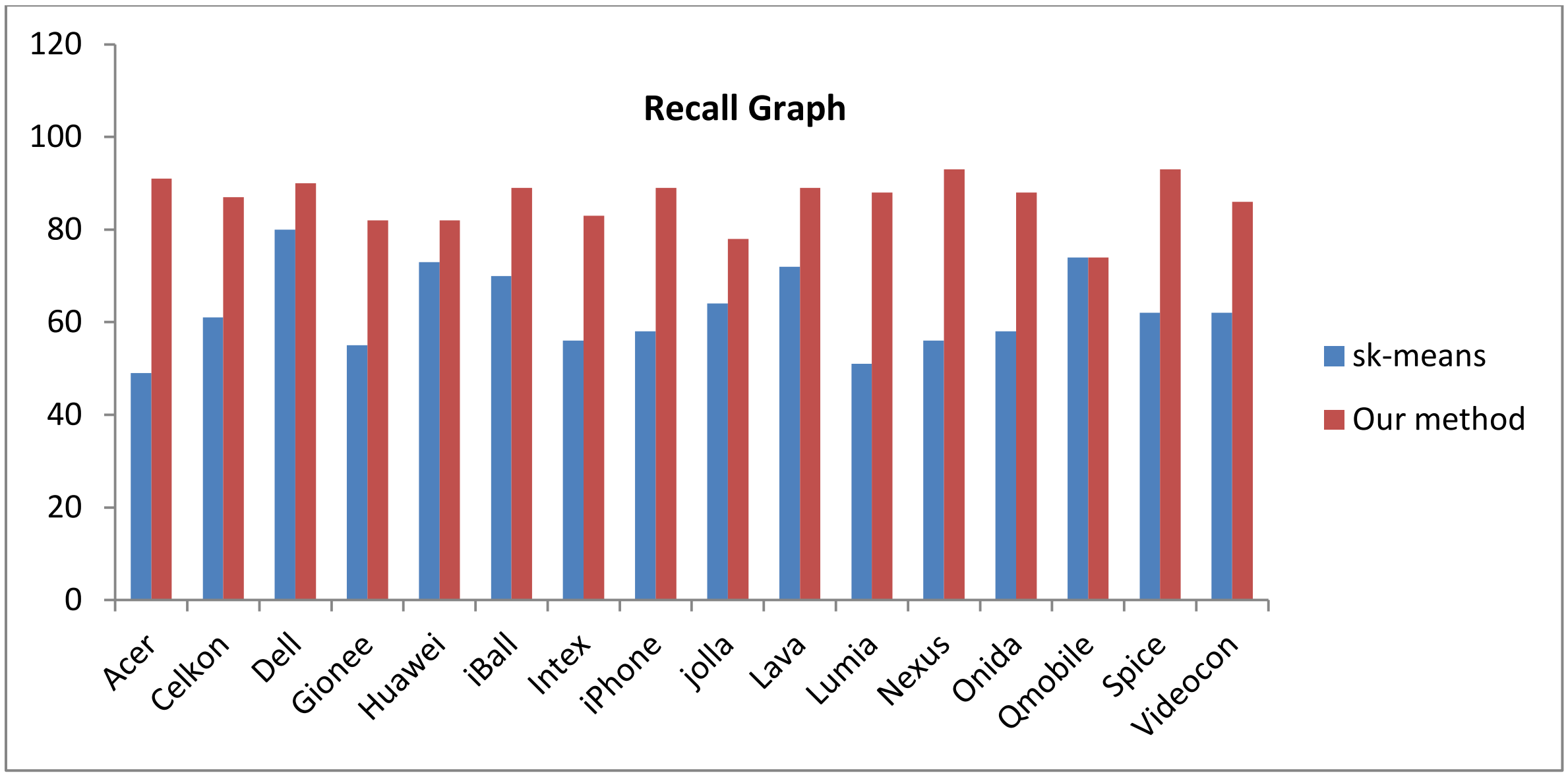

Figure 8: Recall values gained by our method and sk-means

For our 1028 web page dataset, we compared our method with sk-means. Spherical k-means has failed to find all the 27 clusters. It constructed 27 clusters as we provide the number of clusters to it. Only 16 clusters found to have some meaning. Other remaining clusters have mix class distributions. Our method reached high precision and recall values for each cluster.

## 7. Conclusion

Clustering of web search result is a trending topic of research. In this paper we have introduced a method to cluster search result based on enhanced k-means using a heuristic approach. The significant advantage of our approach is that there is no need to specify the factor k. The value of k is decided by heuristic we apply. The heuristic says that some related documents in search result are connected by hyperlinks. We extracted those hyperlinked documents to create initial centroids. In the centroid creation phase only, almost one third cluster members get assigned to the centroids. The proposed method was applied on the live result returned by the Google search engine on the query “jaguar” and “puma”. The result showed a great purity, entropy and precision values. As our future study, we are trying to label the clusters. We further evaluated our method on dataset of web pages. It correctly constructed the 27 clusters. The high quality of clusters does not guarantee the selection by user if the labels are not accurate. Labeling of cluster is an important issue which is being investigated.